\documentclass[sigconf, nonacm, screen]{acmart}

\usepackage{algorithm}
\usepackage{algpseudocode}
\usepackage{bbm}
\usepackage{multirow}

\AtBeginDocument{%
  }

\setcopyright{acmcopyright}
\copyrightyear{2018}
\acmYear{2018}
\acmDOI{XXXXXXX.XXXXXXX}

\acmConference[Conference acronym 'XX]{Make sure to enter the correct
  conference title from your rights confirmation emai}{June 03--05,
  2018}{Woodstock, NY}
\acmPrice{15.00}
\acmISBN{978-1-4503-XXXX-X/18/06}

\begin{document}

\title{Protecting User Privacy in Remote Conversational Systems: A Privacy-Preserving framework based on text sanitization }

\author{Zhigang Kan}
\affiliation{%
  \institution{National University of Defense Technology}
  \city{Changsha}
  \state{Hunan}
  \country{China}
}
\email{kanzhigang13@nudt.edu.cn}

\author{Linbo Qiao}
\affiliation{%
  \institution{National University of Defense Technology}
  \city{Changsha}
  \state{Hunan}
  \country{China}
}
\email{qiao.linbo@nudt.edu.cn}

\author{Hao Yu}
\affiliation{%
  \institution{Technical University of Munich}
  \city{Munich}
  \state{Bavaria}
  \country{Germany}}
\email{hao.yu@tum.de}

\author{Liwen Peng}
\affiliation{%
  \institution{National University of Defense Technology}
  \city{Changsha}
  \state{Hunan}
  \country{China}
}
\email{pengliwen13@nudt.edu.cn}

\author{Yifu Gao}
\affiliation{%
  \institution{National University of Defense Technology}
  \city{Changsha}
  \state{Hunan}
  \country{China}
}
\email{gaoyifu@nudt.edu.cn}

\author{Dongsheng Li*}
\affiliation{%
  \institution{National University of Defense Technology}
 \city{Changsha}
 \state{Hunan}
 \country{China}}
\email{dsli@nudt.edu.cn}

\renewcommand{\shortauthors}{Kan et al.}

\begin{abstract}

Large Language Models (LLMs) are gaining increasing attention due to their exceptional performance across numerous tasks. As a result, the general public utilize them as an influential tool for boosting their productivity while natural language processing researchers endeavor to employ them in solving existing or new research problems. Unfortunately, individuals can only access such powerful AIs through APIs, which ultimately leads to the transmission of raw data to the models' providers and increases the possibility of privacy data leakage. Current privacy-preserving methods for cloud-deployed language models aim to protect privacy information in the pre-training dataset or during the model training phase. However, they do not meet the specific challenges presented by the remote access approach of new large-scale language models.

This paper introduces a novel task, ``User Privacy Protection for Dialogue Models,'' which aims to safeguard sensitive user information from any possible disclosure while conversing with chatbots. We also present an evaluation scheme for this task, which covers evaluation metrics for privacy protection, data availability, and resistance to simulation attacks. Moreover, we propose the first framework for this task, namely privacy protection through text sanitization. Before sending the input to remote large models, it filters out the sensitive information, using several rounds of text sanitization based on privacy types that users define. Upon receiving responses from the larger model, our framework automatically restores privacy to ensure that the conversation goes smoothly, without intervention from the privacy filter.  Experiments based on real-world datasets demonstrate the efficacy of our privacy-preserving approach against eavesdropping from potential attackers.

\end{abstract}

\ccsdesc[500]{Computing methodologies~Information extraction}

\keywords{large language model, privacy-preservation, natural language processing, in-context learning}


\maketitle

\begin{figure}[tb]
  \centering
  \includegraphics[width=\linewidth]{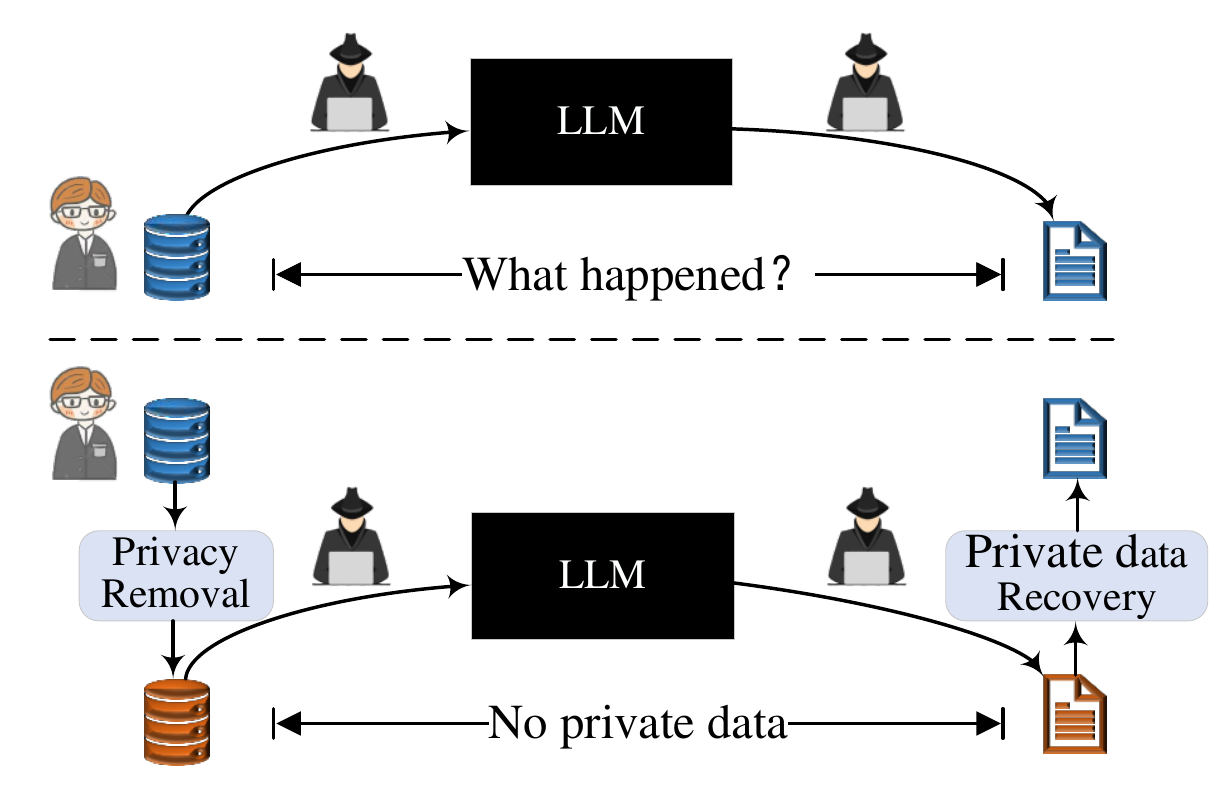}
  \caption{Privacy breach risk faced by remotely accessing large models and a corresponding countermeasure.}
  \label{fig:intro_fig}
\end{figure}

\section{Introduction}

Large models, represented by ChatGPT, have demonstrated remarkable capabilities in natural language processing, including but not limited to text understanding and generation. Nevertheless, due to their massive number of parameters, deploying these models locally requires high-performance hardware facilities with costly maintenance expenses. 
Furthermore, numerous expansive and sophisticated models, including the state-of-the-art GPT-4 \footnote{https://openai.com/gpt-4}, are inaccessible in the open-source domain. 
Consequently, even if users possess the necessary financial resources, they are unable to leverage these exceptional models for local usage.
Fortunately, the OpenAI developers, acting as representatives, have provided Application Programming Interfaces (APIs) that allow users to remotely invoke models with astonishing performance, but with huge parameter quantities. Within this context, users send input text to companies that offer large-scale model services, instead of feeding to local models. These companies feed the input to the artificial intelligence model to obtain output, which is then returned to the user. However, as illustrated in Figure \ref{fig:intro_fig}, utilizing remote models increases the risk of privacy data leaks to potential network eavesdroppers and remote service providers. Thus, while providing users with the opportunity to use large models, this outsourced computing method also brings forth new challenges with regard to safeguarding privacy information.

Current privacy-preserving techniques for language models deployed in the cloud either prevent untrustworthy customers from obtaining privacy in pre-trained datasets \cite{model_attack_01, model_attack_02, model_attack_03} or shield pre-training and fine-tuning datasets of users from untrustworthy cloud service providers \cite{token_leakage, token_leakage_survey}. However, these approaches encounter significant difficulties when faced with the distinct challenges presented by the remote accessibility paradigm of large-scale models.
Traditional privacy protection techniques \cite{Untrusted_service_provider_01, Untrusted_service_provider_02,Untrusted_consumers_01, Untrusted_consumers_02} for cloud computing are inadequate to address the aforementioned emerging challenges, as large model invocations differ from traditional cloud computing scenarios in the three aspects: privacy data presentation, requirements for privacy removal and cloud service mode.

\textit{Privacy data presentation.} Traditional cloud computing privacy protection methods are oriented towards large-scale structured data, protecting the values of certain attributes from potential attackers. These values are known and easy to locate. In the process of invoking remote large models, the data stream from users is often singular and unstructured. Due to the diversity of natural language expression, privacy information that needs to be protected in this process appears in different forms and positions without any regularity in unstructured data. This presents a challenge for privacy identification and location, which traditional data anonymization (k-anonymity, I-diversification), differential privacy and other methods fail to consider.

\textit{Requirements for privacy removal.} Traditional privacy protection methods aim to remove specified types of privacy from a structured data collection while maintaining the overall availability of the data collection. Therefore, they generalize privacy information into fuzzy values through methods such as adding noise, so as to interfere with potential attackers' access to privacy in the data set. However, in scenarios where remote large models are used, the protected object is a single unstructured data, privacy protection methods not only need to completely remove the specified privacy information from the data but also need to  avoid the loss of critical information. Specifically, the resulting unstructured data after privacy removal not only needs to be semantically coherent but also needs to guide the large model to achieve the user's original intent.

\textit{Cloud service mode.} In the context of traditional cloud computing, data providers can segment privacy in data into multiple secret shares, each of which cannot expose the privacy of the original data, but can reconstruct privacy when combined.
Based on the secret sharing scheme, data providers can coordinate multiple cloud service providers to jointly complete tasks through secure multi-party computation technology. However, due to the non-public nature of large model weights and techniques, specific large model services usually have only one provider. Moreover, based on instruction learning, generating text through large models is an end-to-end process, and intermediate values are difficult to share between different large models providers. Therefore, it is difficult to use traditional secure multi-party computation methods to protect privacy information in data during the remote use of large models.

As large-scale models in the field of artificial intelligence continuously achieve astounding breakthroughs, an increasing number of individuals exhibit a strong interest in these AI technologies. This has led to a surge in remote interactions with these large models through the internet. 
However, the current interactive mode faces a dilemma. On the one hand, remote large-scale models require plain text input. On the other hand, users are concerned that sending raw text directly may compromise their privacy and expose them to potential attackers.
Despite the widespread use of remote access to non-open-source large models, there remains an open question about effectively preserving user privacy during interactions with remote artificial intelligence. Therefore, we propose a new task: User Privacy Protection for Dialogue Models, aimed at safeguarding user sensitive data during using remote large models. To this end, we establish the first benchmark for this task and introduce evaluation metrics from two aspects: privacy preservation and data utility.

In this paper, we propose the first privacy protection framework, namely Privacy-Preserving via Text Sanitization (PP-TS), to defend against potential eavesdropping attacks during remotely chatting with large-scale models. This framework applies multiple cycles of text sanitization to de-identify the sensitive data before transmitting the request to the remote server. Upon receiving the response from the remote LLM, it automatically re-identifies and restores the original privacy information, ensuring the privacy protection without impacting the user's experience in model utilization.
The framework is composed of a pre-processing privacy protect module that performs de-identification, a LLM invocation module, and a post-processing privacy recovery module that restores the original sensitive information.
Instruction tuning is adopted to identify and substitute sensitive user data, and contextual learning is introduced to enhance the anonymization capability and text rewriting ability of this module. 
The remote LLM invocation module utilizes API to achieve feedback from the remote large models. 
Finally, the post-processing information restoration module adds private information to the feedback by employing a locally saved sensitive data mapping.

We evaluated our approach for event extraction, a classic task in the field of information extraction. We quantified the performance of our proposed framework using two metrics- availability and anonymization rate. We also simulated attacks from potentially untrusted service providers and reported the ability of our framework to withstand privacy infringement attacks.

In summary, this work makes the following contributions: 
\begin{itemize}
    \item This paper proposes a novel task aimed at protecting privacy information of users during remote invocation of large models. The privacy protection issues in the remote large model invocation process are completely different from traditional privacy protection tasks for cloud computing, but they are urgently in need of resolution.
    \item A new evaluation scheme is introduced for this task, which includes quantifying the performance of privacy protection methods from two aspects: privacy preservation and data utility, as well as reporting the performance of privacy protection methods under simulated attacks.
    \item We introduce the first user privacy information protection framework for remote invocation of large models. It effectively prevents specified types of privacy through a pre-processing data de-identification module and a post-processing privacy recovery module.
\end{itemize}

\section{Related Work}

\subsection{Privacy-preservation Technologies}

Cloud computing offers computing services to a large number of users while also harboring the inherent risk of exposing user privacy data. 
The risk stems from two primary sources: untrustworthy cloud service providers and untrustworthy data consumers. 
The former has the potential to pilfer private user data either during the input phase or computing processes \cite{Untrusted_service_provider_01, Untrusted_service_provider_02}. 
The latter can execute malicious programs through the cloud platform's service interface to snoop confidential information by analyzing returned results \cite{Untrusted_consumers_01, Untrusted_consumers_02}.

Various privacy-preservation techniques have been designed by researchers from different perspectives to ensure the confidentiality of user information.
Data separation methods leverage distributed computing \cite{Data_separation_distribution_01, Data_separation_distribution_02, Data_separation_distribution_03, Data_separation_distribution_04, Data_separation_distribution_05} or federated learning \cite{Federated_learning_01, Federated_learning_02, Federated_learning_03, Federated_learning_04, Federated_learning_05, Federated_learning_06} to store and process sensitive data on local or trusted computing nodes, and non-sensitive data on untrusted nodes.
Techniques based on data interference, including data anonymization \cite{data_anonymization_00, data_anonymization_01, data_anonymization_02} and differential privacy \cite{DP_01, DP_02, DP_03, DP_04, DP_05}, impede attackers from accessing accurate private information by introducing noise to the original data.
Secure multiparty computation technologies \cite{SMPC_01, SMPC_02, SMPC_03} divide private information into several secret shares, none of which alone reveal any confidential information about the original data. However, when combined, they can reconstruct private information. 
As a result, data providers can collaborate with multiple cloud service providers to complete tasks using secret sharing schemes while addressing privacy concerns. 
In addition, hardware-based methodologies \cite{intel_01, intel_03} leverage trusted execution environments, such as Intel SGX \cite{intel_sgx} and ARM TrustZone \cite{AMR_t} to safeguard the confidentiality of critical data and code.

\subsection{Privacy Protection for Language Models}

Privacy protection for language models consists of two independent families: protecting personal privacy in the pre-training corpus and protecting sensitive information in user's input data \cite{LLM_privacy_survey}. The former aims to prevent attackers from inducing models to disclose privacy in pre-training data \cite{model_attack_01, model_attack_02, model_attack_03}. Introducing differential privacy \cite{DP_original_paper} and its derivatives \cite{DP_derivatives} to pre-train \cite{DP_derivatives_01} a DP-protected model is a potent means of mitigating this risk \cite{DP_SGD}. Despite the usefulness of DP, the incorporation of differential privacy may result in a decrease in the model's overall utility. As such, some scholars propose alternate techniques such as personalized DP \cite{personal_DP}, one-sided DP \cite{one_side_DP}, and Selective-DP \cite{selective_DP} to augment the model's utility. Alternatively, modifying the model's training process \cite{training_process_DP_01, training_process_DP_02, training_process_DP_03} or structural design \cite{structural_design_DP} is another feasible strategy to incorporate differential privacy while simultaneously preserving the model's utility. 
Recently, with the emergence of generative large-scale language models such as GTP3, scholars proposed DP-based large model-oriented pre-training \cite{DP_LLM_pre_train}, fine-tuning \cite{finetuning_DP} and inference \cite{DP_LLM_inference} techniques as a means of addressing potential private leaking risks.

Protecting sensitive information in user input data primarily concerns addressing the privacy risks that users encounter when uploading data for training or pre-training models in the cloud \cite{token_leakage, token_leakage_survey}. An uncomplicated strategy to mitigate this issue involves transforming user data into unreadable token representations and thereafter perturbing them using differential privacy techniques before uploading them to the cloud \cite{token_denf_01, token_representation_def_01, token_representation_def_02, token_representation_def_04, token_representation_def_05, token_representation_def_06}. Nonetheless, adversaries can still utilize text reconstruction techniques \cite{Information_Leakage_01, Information_Leakage_02, Information_Leakage_03} to restore the original information from token representations. To tackle this challenge, Feng et al. \cite{homomorphic_encryption_01} and Chen et al. \cite{The_X} sequentially proposed theoretically guaranteed approaches to ensuring privacy by applying homomorphic encryption techniques to encrypt both user input data and computation processes. Alternatively, scholars try to safeguard user input at the source by substituting private data with non-sensitive tokens \cite{UMLDP, token_replace_02}. However, this will result in a breakdown of semantic consistency in the text \cite{token_fusion_01}. Contrarily, token reduction \cite{token_reduction_01, token_reduction_02, token_reduction_03} and fusion \cite{token_fusion_01} are two efficacious methodologies for thwarting text recovery attacks. These techniques curtail or merge extraneous token representations during the encoding process, thus impeding attackers from obtaining sensitive information through the reconstructed text.

\subsection{Text Sanitization}

Research in the area of text sanitization can be broadly classified into two categories: one involves the identification and removal of sensitive information from data using NLP techniques, while the other employs Privacy Preserving Data Disclosure (PPDP) techniques \cite{Text_Sanitization_survey_01}. 
In the realm of NLP, text anonymization is viewed as a problem of Named Entity Recognition (NER). 
Early approaches utilized rule-based methodologies to cleanse pre-defined categories of sensitive information \cite{rule_Text_Sanitization_01, rule_Text_Sanitization_02}. 
However, with the advent of advanced neural network techniques, the mainstream shifted towards employing manually labeled data to train models to automate the sanitization process \cite{NN_Text_Sanitization_01, NN_Text_Sanitization_02, NN_Text_Sanitization_03}.
PPDP approaches \cite{PPDP_01, PPDP_02, PPDP_03} formulate the problem of anonymizing text as seeking for the minimum information set of masking operations, such as data suppression or generalization, to impose compliance with the requisites emanating from the privacy framework.

\begin{figure*}[thb]
  \centering
  \includegraphics[width=\linewidth]{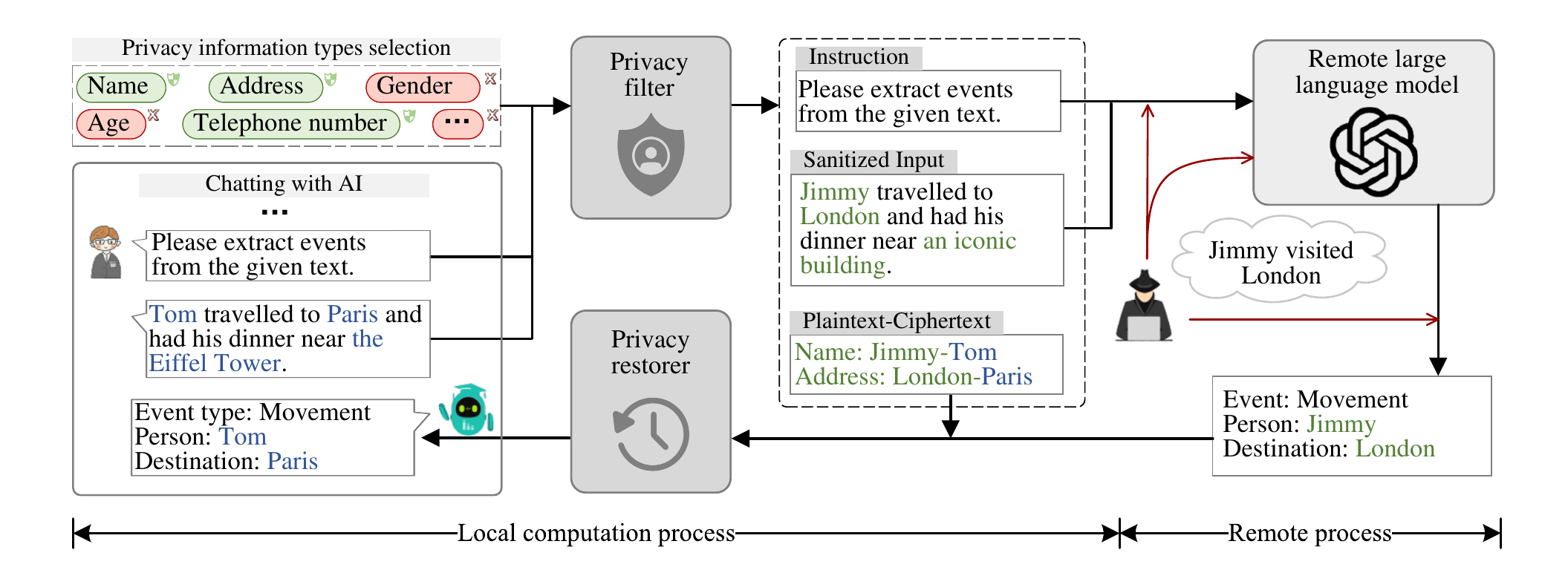}
  \caption{The workflow for protecting user's sensitive information during chatting with remote dialogue models via our text sanitization based privacy-preserving framework.}
  \label{fig:overview}
\end{figure*}

\section{Task Definition}

\subsection{Problem Formulation}

For a given user $U$ who wishes to utilize a remote, non-open-sourced large model $L$ without compromising sensitive information across various aspects ${A_1, A_2, \cdots, A_n}$, where $A_i$ represents the i-th privacy attribute (e.g. name or age). 
The user's input $X$ may contain several pieces of private information $P={P_{A_1}, P_{A_2}, \cdots, P_{A_n}}$, where $P_{A_i}={P_{A_i}^1, P_{A_i}^2, \cdots, P_{A_i}^m}$ is the collection of private information pertaining to attribute $A_i$, and $P_{A_i}^j$ represents the $j$-th piece of private information pertaining to attribute $A_i$. 
Based on input $X$, the output $Y$ of the large model $L$ includes the expected information $I$. The aim of privacy protection is to achieve the following objectives during the interaction between user $U$ and the remote large model $L$: 1) adding disturbance to $X$ to obtain $\hat{X}$, making it impossible for potential attackers to directly acquire or infer any private information from $P$ in $\hat{X}$, 2) using $\hat{X}$ as input for large model $L$ to generate output $\hat{Y}$. The privacy protection method can recover the expected information $I$ from $\hat{Y}$.

\subsection{Evaluation Scheme}

On the one hand, during chatting with remote large-scale models, preserving privacy methods should prevent potential attackers from obtaining private information.
On the other hand, they should retain sufficient usable information for guiding the large-scale models to achieve the original intention of users.
Therefore, evaluating the performance of privacy protection techniques for unstructured data requires measuring their abilities in two aspects: 1) privacy protection, and 2) data utility.
Furthermore, despite the interference of privacy protection methods on sensitive information in unstructured data, potential attackers may still infer useful information from the general semantic context of the data.
Take the input ``I live in Paris and I can see the Eiffel Tower from my house'' as an example.
Even if a privacy protection algorithm replaces the user's residential city ``Paris'' with ``London'', acquiring the aforementioned ``privacy protection'' capacity, a potential adversary can still deduce that the real location of the user is either ``Paris'' or ``London'' based on the text ``I can see the Eiffel Tower from my house''. 
Therefore, this paper also advocates for quantifying the performance of privacy protection methods through simulation of privacy attacks.

\subsubsection{Privacy preservation}

In terms of privacy preservation, we are committed to reporting the ability of privacy protection methods to identify, locate, and interfere with confidential information.
We employ the Privacy Removal Rate (PRR) to assess the efficacy of methods in safeguarding privacy. 
Given a input set $X={x_1, x_2, ..., x_s}$ whereby $x_i$ denotes the $i$-th input containing sensitive information $P^{X_i}={P^{x_i}_{A_1},P^{x_i}_{A_2}, \cdots, P^{x_i}_{A_n}}$, the processed input set $\hat{X}={\hat{x_1}, \hat{x_2}, \cdots, \hat{x_s}}$ entails sensitive data $\hat{P}^{x_i}={\hat{P}^{x_i}_{A_1},\hat{P}^{x_i}_{A_2}, \cdots, \hat{P}^{x_i}_{A_n}}$ after undergoing privacy protection algorithms. The privacy removal ratio can be represented as:

\begin{equation}
   \text{PRR} = 1 - \frac{\sum_{i=1}^{s} \sum_{j=1}^{n} \text{card}(\hat{P}^{x_i}_{A_j})}{\sum_{i=1}^{s} \sum_{j=1}^{n} \text{card}(P^{x_i}_{A_j})} ,
\end{equation}
where $\text{card}(\cdot)$ indicates the cardinal number of the corresponding set.

\subsubsection{Data utility}

We quantify the capacity of privacy preservation methods in retaining useful information through the metric of Data Utility Rate (DUR).
Specifically, this rate is defined as the proportion of processed data that can guide the large model to attain the users' primary intentions.
Given the original input set $X$, the output of the model without any privacy protection process is $Y={Y_1, Y_2, \cdots, Y_s}$, while the information set that users expect to obtain is $I={I_1, I_2, \cdots, I_s}$.
By utilizing privacy protection methods, the input set of the model $L$ is transformed to $\hat{X}$, and the output set is $\hat{Y}= {\hat{Y}_1, \hat{Y}_2, \cdots, \hat{Y}_s}$. 
The information set obtained from $\hat{Y}$ after privacy data recovery is represented as $\hat{I}= {\hat{I}_1, \hat{I}_2, \cdots, \hat{I}_s}$. 
The formalization of data utility rate can be expressed as:

\begin{equation}
   \text{DUR} = \frac{\sum_{i=1}^{s} \mathbbm{I}(I_i=\hat{I}_i) }{\text{card}(X)} ,
\end{equation}
where $\mathbbm{I}(\cdot)$ is the indicator function.

\subsubsection{Resisting simulated attacks}


The present paper proposes a methodology to evaluate the practicality of privacy-preserving methods via simulated attacks, and introduces the Data Protection Rate (DPR) as the evaluation metric. 
Specifically, this rate refers to the proportion of processed data that accurately guides a model to achieve the user's initial objective, while preventing attackers from inferring any private information.

\begin{equation}
   \text{DPR} = \frac{\sum_{i=1}^{s} \mathbbm{I}(I_i=\hat{I}_i) * \mathbbm{I}(\text{Attack unsuccessful}) }{\text{card}(X)} .
\end{equation}

\section{Privacy-preserving via Text Sanitization}

\subsection{The Privacy-preserving Framework for Applying Generative Language Models}

The paper presents a privacy-preserving framework for chat scenarios with conversational large language models. Figure \ref{fig:overview} illustrates the framework's workflow, which comprises of three modules: privacy protecting, large model inference, and privacy recovery. The privacy protecting module and the privacy recovery module are deployed locally, while the large model inference module utilizes APIs to access large-scale language models in the cloud. To begin, users are prompted to choose the type of privacy they wish to protect. During the dialogue, the privacy protecting module automatically filters sensitive information from users' input, while retaining the plaintext-ciphertext set. The large model inference module generates responses based on the sanitized input via APIs. The privacy recovery module then accepts the model's output, recovers the privacy in the response based on the plaintext-ciphertext set kept locally, and provides feedback to the user.

\begin{figure}[tb]
  \centering
  \includegraphics[width=\linewidth]{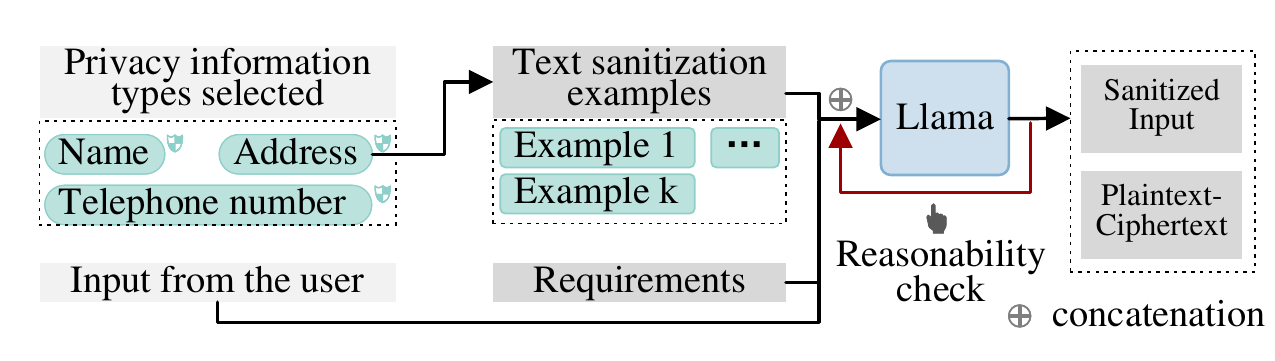}
  \caption{The process of text sanitization with a pre-determined type for the user input.}
  \label{fig:privacy_protector}
\end{figure}

\subsection{Privacy Protecting Module}

In our framework, the privacy protecting module dedicated to ensuring privacy is a filter that sits between the user and the remote model. This module allows users the flexibility to choose the privacy type that may require sheltering. The protector receives user input, identifies and eliminates any sensitive data and subsequently forwards it to the language model present on the cloud. In this paper, our privacy protector transforms privacy filtering into a specified privacy content replacing and multiple text rewriting task.
To accommodate for varying privacy types that may have been chosen by the user, the module filter out the privacy information by repeatedly sanitize the user input in a cyclic manner. Algorithm \ref{alg:filter} elucidates the full user-input privacy-filtering process by the privacy protection module. Each loop iteration entails one data sanitization process for a specific type $A_i$. Figure \ref{fig:privacy_protector} portrays the flow of one data sanitization cycle.

With ``Address'' as the user-selected privacy type, the Text Protector retrieves ``Rewrite Requirements'' and ``Text Sanitization Examples'' instructions from a pre-designed library without any human intervention. 
Afterward, the ``Rewrite Requirements,'' ``Text Sanitization Examples,'' and user input are combined and fed into a generative language model that is decoder-only, with substantially fewer parameters than ChatGPT.
Eventually, the generative model outputs the rewritten input and the Plaintext-Ciphertext record based on the ``Rewrite Requirements'' and ``Text Sanitization Examples.'' For instance, when presented with input such as ``\textit{Tom travelled to Paris and had his dinner near the Eiffel Tower},'' the privacy filter may replace the location ``\textit{Paris}'' with ``\textit{London}'' at random, resulting in an originally sanitized input text.

Replacing privacy information in the text can lead to semantic inconsistency for the sanitized input, as seen in the example above where the introduction of ``\textit{London}'' contradicts ``\textit{the Eiffel Tower}'' in the original input text. This inconsistency not only causes issues for remote large models to understand the input, but also presents an opportunity for potential attackers to gain access to genuine privacy information. 
To mitigate this situation, this research presents a mechanism to verify and rectify the rephrased text's plausibility in the text sanitization process, which is demonstrated in lines 7 to 13 of Algorithm \ref{alg:filter}. The privacy-preserving module identifies the rewritten text X's plausibility by using a local deployed generative model based on instruction learning.
If X is semantically contradictory, the content (``\textit{the Eiffel Tower}'') in the original text that conflicts with the rewritten part (``\textit{London}'') is rewritten using a local model based on contextual learning.
For instance, ``the Eiffel Tower'' is rephrased as the more abstract description of ``an iconic building.''

\begin{algorithm}
\renewcommand{\algorithmicrequire}{\textbf{Input:}}
\renewcommand{\algorithmicensure}{\textbf{Output:}}
\caption{Filtering private information}\label{alg:filter}
\begin{algorithmic}[1]
\Require User Input $X$, privacy types $A$,
\Ensure Sanitized input $\hat{X}$, plaintext-ciphertext set $PCS$;
\State Initiate $\hat{X} \gets X$\;
\For{$A_i$ in $A$}
    \State Construct text sanitization requirements $R$ for the type $A$\;
    \State Construct text sanitization examples $E$ for the type $A$\;
    \State $I_p \gets  R \oplus E \oplus \hat{X}$\;
    \State Feed $I_p$ into Llama to obtain the sanitized text $\hat{X}$ and the Plaintext-Ciphertext record $pcr$\;
    \State Do a reasonability check on $\hat{X}$\;
    
    \While{$\hat{X}$ is contradictory}
        \State Fix inconsistencies in sanitized text $\hat{X}$\;
        \State Do a reasonability check on $\hat{X}$\;
        \If{$\hat{X}$ is contradictory}
            \State Feed $I_p$ into Llama to obtain a new $\hat{X}$ and $pcr$\;
        \EndIf
    \EndWhile
    
    \State Append $pcr$ to $PCS$\;
\EndFor
\end{algorithmic}
\end{algorithm}

\subsection{The Large Model Inference and Privacy Recovery Modules}

The large model inference module utilizes APIs to access remote large-scale language models and takes sanitized texts as inputs. The privacy recovery module is specifically designed to receive responses from cloud models and to automatically recover the privacy that was initially obscured in the privacy protection module. This module replaces the corresponding part of the large model response with privacy content, which is based on the ciphertext-original set that was locally stored during the privacy filtering process, via In-context learning.

\begin{table*}[thb]
\centering
\renewcommand\arraystretch{1}
\caption{The performance of our privacy-preserving approach against simulated attacks (\%).}
\label{table:main_res_s_attack}
\setlength{\tabcolsep}{2.5mm}{
\begin{tabular}{lccccc}
\toprule
\hline
\multirow{2}{*}{Method} & \multicolumn{2}{c}{Manual attack}     &   ~    & \multicolumn{2}{c}{Procedural attack}  \\
\cline{2-6}

      & Literal detection DPR    & logical inference DPR   &   ~    & Literal detection DPR    & logical inference DPR     \\ \toprule
                       
PP-TS     & 93.00 & 89.00   &   ~   & 94.33 & 91.00 \\
~-~ \textit{w/o} reasonability check   & 77.00 & 70.00  &   ~   & 79.00 & 74.67  \\
~-~ \textit{w/o} Privacy filter  & 0.00 & 0.00 &   ~   & 11.00 & 19.00   \\
\hline
\bottomrule
\end{tabular}}
\end{table*}

\section{Experiment}

\subsection{Experimental Setups}

\subsubsection{Implementation Details}

To visually present the efficacy of our proposed approach for filtering and recovering different types of privacy, we anticipate that users will regularly require various types of information during their conversations with large models. Consequently, we evaluate our proposed approach on the event extraction task in this paper. This chapter assumes that the user is carrying out event extraction via conversation with a remote large model while keeping their dataset confidential. To safeguard the dataset from being divulged, our privacy-preserving approach is implemented in the process.

In this work, we employ the existing large-language-model-based approach \cite{ChatGPT_base_EE_01} for event extraction to recreate the conversational interaction between AI systems and users. Specifically, we leverage Llama-7B \cite{Llama} (a local generative language model) and ChatGPT\footnote{https://openai.com/blog/chatgpt} (gpt-3.5-turbo version) (a remote big-predicate model) for this purpose.
We select the ACE2005 as the evaluation dataset, which is the most frequently employed event extraction dataset by many previous authors. We randomly sample 100 entries from it to construct our test set.

\subsubsection{Simulated Attack}

In this paper, we simulate a potential attack on user privacy by identifying a specified type of information directly or inferring them from sanitized input. Specifically, this paper sets name and location as the types of privacy to be protected in the dataset.
To accomplish this objective, we use both manual and procedural approaches to extract specified types of privacy information from sanitized text. For the manual approach, we invited three graduate students with NLP research experience to independently count the privacy information of each type in the test set. In this paper, the average of their results is taken as the final result. We employ a ChatGPT-based (gpt-3.5-turbo version) privacy detector as the procedural approach, which utilizes in-context learning to identify or deduce the privacy information with specified types in user input.

\subsubsection{Evaluation Metric}

 In this chapter, we shall employ the three metrics introduced in Section 3.2 to evaluate the pragmatic efficacy of our privacy-preserving approach. 
 Simulation attacks can be classified into two levels: the first is directly detecting a specific type of privacy literally, while the second involves the extraction of private details by delving into other contents present within the text. Our nomenclature system defines the former as ``literal detection'', while the latter is named as ``logical inference''.
 The Data Protection Rate (DPR) at the ``literal detection'' level is conceptually similar to the Privacy Removal Rate (PRR), but it is completely different. PRR only requires a count of obscured literal privacy, while DPR takes into account the adequacy of modified data as well.
 In addition, follow the previous work, we employ the P (Precision), R (Recall), and F1 (F1-Score) measures to demonstrate the event extraction performance.

\subsection{Main Result}

\begin{table}[thb]
\centering
\renewcommand\arraystretch{1}
\caption{The result of privacy-preserving on the event extraction task (\%).}
\label{table:main_res}
\setlength{\tabcolsep}{4mm}{
\begin{tabular}{lcc}
\toprule
\hline

Method      & PRR    & DUR        \\ \toprule
                       
PP-TS     & 95.96 & 92.33   \\
~-~ \textit{w/o} reasonability check   & 95.96 & 81.67  \\
~-~ \textit{w/o} Privacy filter & 0 & 100   \\


\hline
\bottomrule
\end{tabular}}
\end{table}

Table \ref{table:main_res_s_attack} visualizes the performance of the privacy-preserving approach proposed in this paper in resisting simulation attacks, and Table \ref{table:main_res} demonstrates the approach's ability to filter privacy and retain data availability.

From Table \ref{table:main_res_s_attack}, we can observe that PP-TS can effectively protect the privacy information in user data regardless of manual or procedural attacks. 
Compared to literal detection, this method exhibits lower resistance to attacks at the logical inference level, indicating that privacy information might be reflected in deep semantics within the given dataset in addition to the literal representation. 
After the reasonability check mechanism is removed from the sanitized text, PP-TS displays a significantly lower resistance to attacks, demonstrating that adjusting the semantic contradiction of the sanitized text is a useful solution in preventing potential attackers from drawing conclusions about specific privacy information from the overall semantics of the text.
In comparison to manual attacks, PP-TS performs better in defending against procedural attacks due to programs having lower ability than human beings in identifying and reasoning out privacy.
The results presented in Table \ref{main_res} validate the efficacy of PP-TS in privacy preservation and data usability.
With the removal of the reasonability checking mechanism, PPR remains unaffected whereas the DUR displays significant reduction. This corroborates the effectiveness of the reasonability check mechanism reasonability check mechanism.

\subsection{Discussion}

\begin{table}[thb]
\centering
\renewcommand\arraystretch{1}
\caption{Performance comparison of event extraction before and after using our privacy-preserving method (\%).}
\label{table:ee_performance}
\setlength{\tabcolsep}{2mm}{
\begin{tabular}{lcccc}
\toprule
\hline
\multirow{2}{*}{Method} & \multicolumn{4}{c}{Trigger}       \\
\cline{2-5}
 & P & R & F1 & $\Delta$F1 \\ \hline
PP-TS     & 57.23 & 42.71 & 48.91 & - \\
~-~ \textit{w/o} reasonability check   & 55.71 & 40.96 & 47.20 & \textcolor{blue}{-1.71} \\ 
~-~ \textit{w/o} Privacy filter & 63.49 & 43.65 & 51.73 & \textcolor{red}{+2.82} \\ 
\hline
 & \multicolumn{4}{c}{Argument} \\ \cline{2-5}
  & P & R & F1 & $\Delta$F1 \\ \hline
PP-TS   & 22.49 & 19.7 & 21.00 & - \\
~-~ \textit{w/o} reasonability check  & 18.64 & 19.41 & 19.02 & \textcolor{blue}{-1.98} \\
~-~ \textit{w/o} Privacy filter  & 25.31 & 21.35 & 23.16 & \textcolor{red}{+2.16} \\
\hline
\bottomrule
\end{tabular}}
\end{table}

The impact of using privacy-preserving methods on the performance of event extraction via chatting with the large model is illustrated in Table 3. While the use of PP-TS for privacy protection may lead to minor degradation in the model's event extraction performance, the impact is negligible overall. The presence of the reasonability check mechanism will improve the performance to some extent.

\section{Conclusion}

Towards the emerging large conversation model, this paper proposes a novel task for the purpose of protecting the privacy of users in conversations. To comprehensively assess privacy-preserving techniques for this task, we present an evaluation scheme that encompasses privacy preservation, data availability, and resistance to simulation attacks. Additionally, we introduce the first framework for this task, which protects privacy through multiple cycles of text sanitization. This framework automatically filters sensitive information from user input through context learning based on user-defined privacy types while also recovering privacy from large model responses, thus eliminating interference with the conversation caused by the privacy filter. The proposed approach is evaluated on the event extraction task, which frequently involves multiple types of information, and experiments demonstrate its effectiveness in defending against potential attacks.


\bibliographystyle{ACM-Reference-Format}
\bibliography{sample-base}

\end{document}